\documentclass[usenatbib,conference]{basi}
\usepackage[T1]{fontenc}
\usepackage[british]{babel}
\usepackage[varg]{txfonts}
\usepackage{rotating}
\usepackage{dcolumn}
\newcommand{\apj}{ApJ}
\newcommand{\apjl}{ApJ Letters}
\newcommand{\apjs}{ApJ Supp.}
\newcommand{\aj}{AJ}
\newcommand{\aap}{A\&A}
\newcommand{\mnras}{MNRAS}
\newcommand{\pasp}{PASP}

\begin{document}
\title[SpeX Prism Library]{The SpeX Prism Library: 1000+ Low-resolution, Near-infrared Spectra of Ultracool M, L, T and Y Dwarfs}
\author[A.~J.~Burgasser]%
       {Adam~J.~Burgasser$^1$\thanks{email: \texttt{aburgasser@ucsd.edu}}\\
       $^1$UC San Diego, 9500 Gilman Drive, La Jolla, CA 92093-0424, USA\\}

\pubyear{2014}
\volume{10}
\pagerange{\pageref{firstpage}--\pageref{lastpage}}

\date{Received --- ; accepted ---}

\maketitle
\label{firstpage}

\begin{abstract}
The SpeX Prism Library (SPL) is a uniform compilation of low-resolution
($\lambda/\Delta\lambda \approx 75-120$), near-infrared (0.8--2.5~$\mu$m)
spectra spanning a decade of observations
with the IRTF SpeX spectrograph.  Primarily containing ultracool M, L, T and Y dwarfs,
this spectral library has been used in over 100 publications to date, facilitating a broad range of
science on low mass stars, exoplanets, high redshift sources and instrument/survey design.
I summarize the contents of the SPL and highlight a few of the key scientific results 
that have made use of this resource, as well as applications in education, outreach and art. 
I also outline the future plans of the SPL, which include 
a reanalysis of early data, better integration and dissemination of source and spectral metadata, 
conversion to Virtual Observatory formats,
development of a Python software package for community analysis, and a design for a node-based
visual programming platform that can facilitate citizen science and project-based learning in 
stellar spectroscopy.\\[6pt]
\hbox to 30pt{\hfil}\verb|http://www.browndwarfs.org/spexprism|
\end{abstract}


\section{SpeX Spectroscopy of Ultracool Dwarfs}\label{s:intro}

The very lowest mass stars and brown dwarfs, collectively referred to as ultracool dwarfs\footnote{We adopt the definition of \citet{1995AJ....109..797K} that an ultracool dwarf has a spectral classification M7 or later, which implies masses $\lesssim$0.1~M$_{\odot}$.} emit the majority of their radiant flux at near-infrared (NIR) wavelengths, so both their discovery and characterization has been facilitated by NIR spectroscopic instrumentation. Notable among these is the SpeX spectrograph on the 3m NASA Infrared Telescope Facility (IRTF; \citealt{2003PASP..115..362R}), an instrument that provides broad-band NIR spectra (0.8-5~$\mu$m) in both multi-order, cross-dispersed, moderate resolution ($\lambda/\Delta\lambda \approx 2000$) and single-order, prism-dispersed, low-resolution ($\lambda/\Delta\lambda \approx 100$) modes.  The latter is well-matched in sensitivity to wide-field red optical and infrared imaging surveys such as 2MASS, DENIS, SDSS, UKIDSS, CFHTLAS, PanSTARRS and WISE, and the spectra of UCDs are easily characterized at low resolution by their broad molecular absorption features. As such, SpeX has been a discovery machine for late M, L, T and Y dwarfs identified in imaging surveys (e.g., \citealt{2004AJ....127.2856B,2006AJ....131.2722C,2008ApJ...676.1281M,2011ApJS..197...19K,2011ApJ...740L..32L}). SpeX spectra have also proven ideal for NIR classification \citep{2006ApJ...637.1067B,2010ApJS..190..100K}, characterization of  atmospheric and physical properties \citep{2006ApJ...639.1095B,2007ApJ...657..511A,2010ApJS..190..100K} and testing atmosphere models \citep{2009ApJ...697..148B,2011A&A...529A..44W}.

\section{The SpeX Prism Library}\label{s:library}

\subsection{Contents and Structure}\label{s:contents}

\begin{figure}
\centerline{\includegraphics[height=3.5cm]{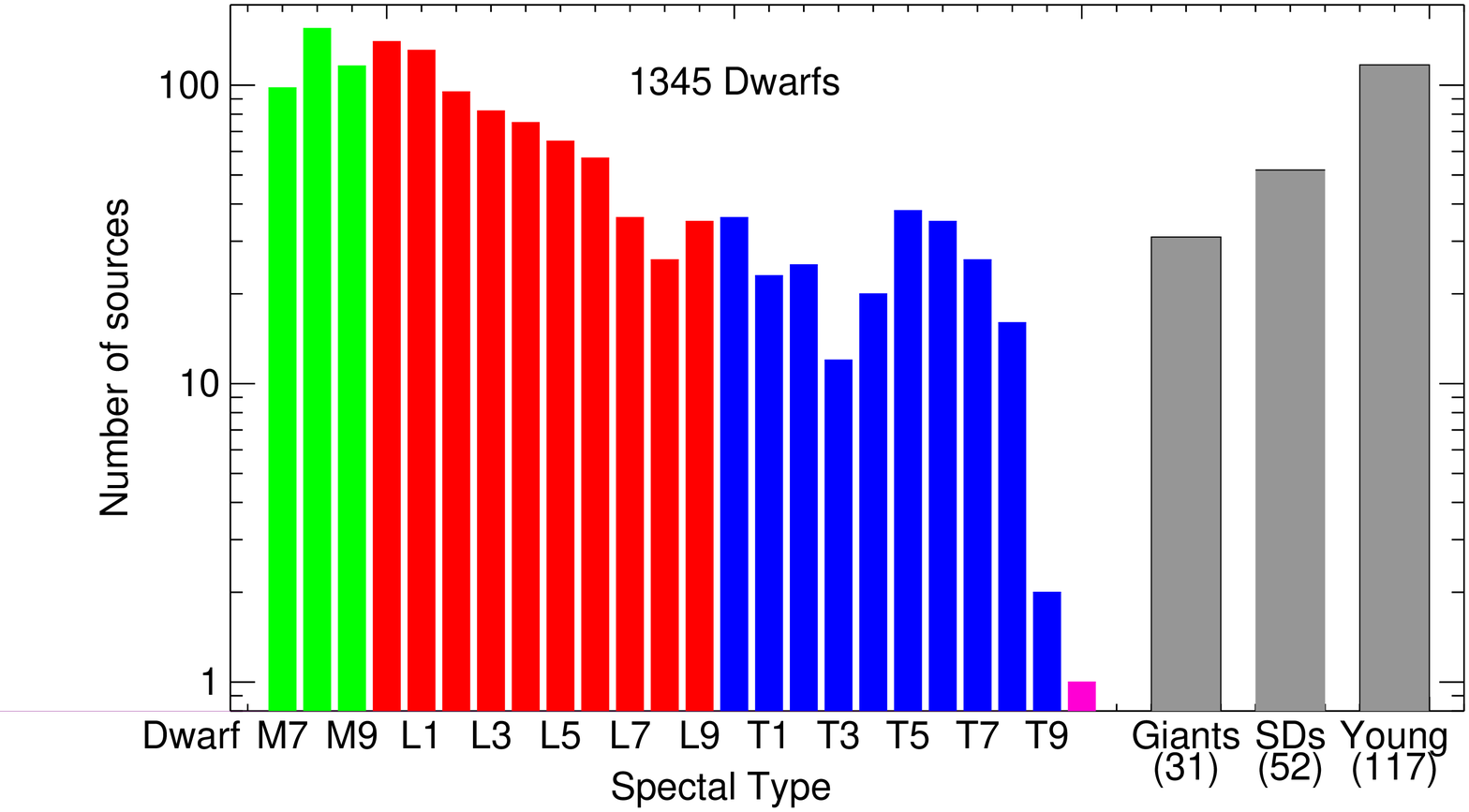}
\includegraphics[height=3.5cm]{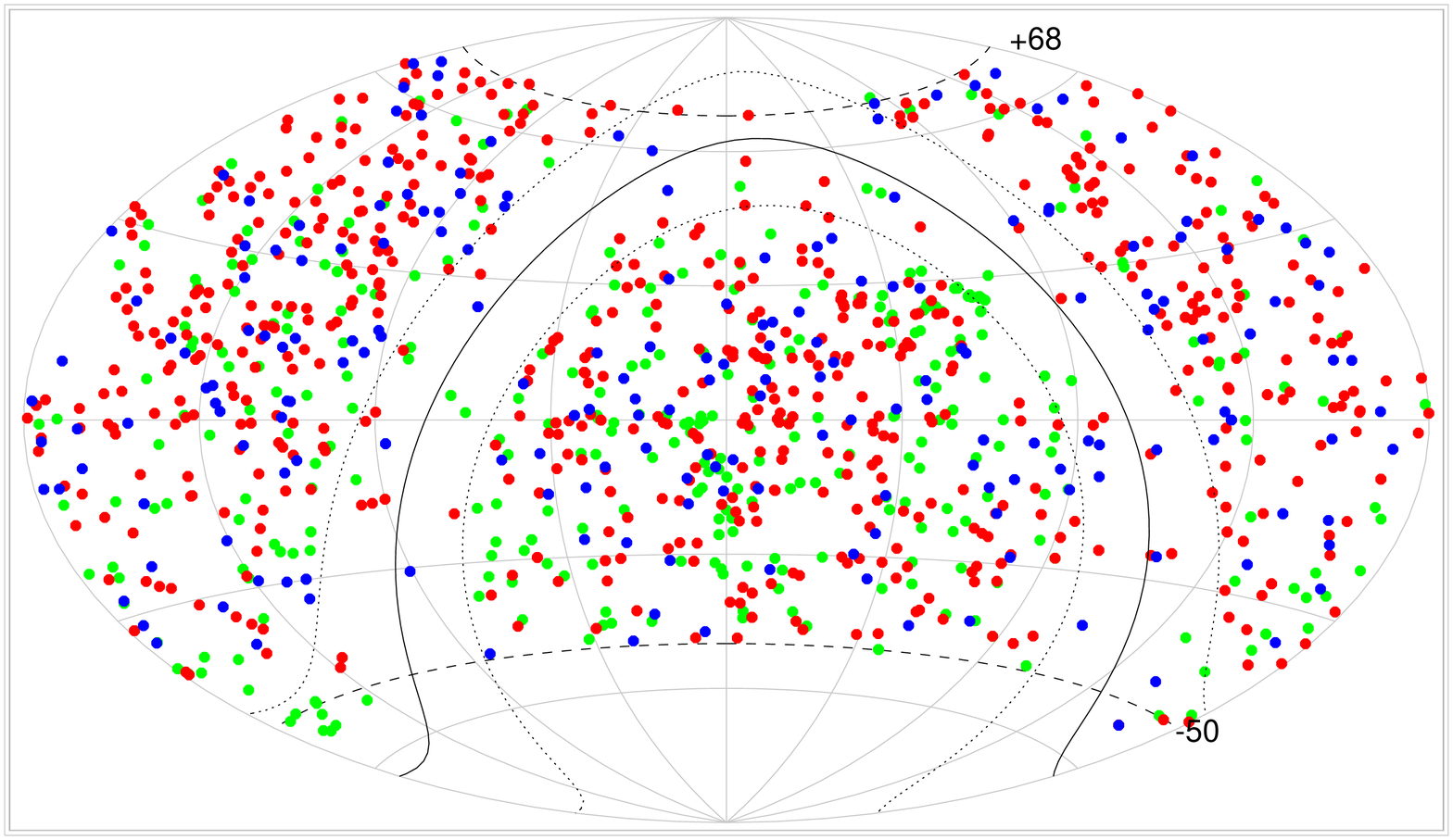}}
\caption{Distributions of SPL spectra by spectral type and class (left) and in equatorial coordinates (right).  UCDs are color coded by spectral type: late-M dwarfs (green), L dwarfs (red), T dwarfs (blue) and Y dwarfs (purple). 
 \label{f:contents}}
\end{figure}

The SpeX Prism Library (SPL) was created in 2008 as a means to organize and disseminate published prism data and facilitate classification of L and T dwarf discoveries using the NIR schemes defined in \citet{2006ApJ...637.1067B} and \citet{2007ApJ...659..655B}.  The initial sample contained  a few hundred spectra of mostly M, L and T dwarfs, all uniformly extracted and calibrated using the SpeXtool reduction package \citep{2003PASP..115..389V,2004PASP..116..362C}.  Since then, the library has grown to over 1900 spectra observed over the past decade, primarily of UCDs ($\approx$1350 spectra; Figure~\ref{f:contents}) but also giant stars, subdwarfs, white dwarfs, carbon stars, novae and supernovae, solar giant planets and galaxies.  A significant fraction of the spectra have been contributed by members of the community.  
The sources span most of the visible sky for IRTF ($-$50$^o$ $\gtrsim \delta \gtrsim$ +68$^o$), with notable gaps around the Galactic plane (poorly sampled in UCD search programs; Figure~\ref{f:contents}). 
The data are of high quality; 50\% (80\%) of the UCD spectra in the SPL have signal-to-noise S/N $>$ 65 (S/N $>$ 30).

\noindent Half of the SPL is currently available on the project website;\footnote{http://www.browndwarfs.org/spexprism} the remainder is scheduled for uploading in early 2014.  The spectra are organized into several libraries based primarily on spectral classes and types. Users can download spectra (wavelength, normalized $f_{\lambda}$ and uncertainty in ascii tables) individually or in batches, and can view ``quicklook'' images of the spectra and source metadata (source positions and designations, spectral types, 2MASS photometry, source and spectral references), but the website does not yet have source filtering or online analysis tools.

\subsection{Science}\label{s:science}

\begin{figure}
\centerline{\includegraphics[height=9cm]{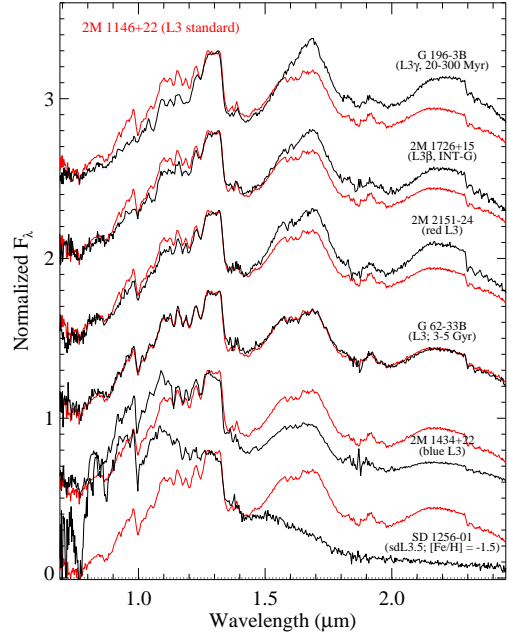}}
\caption{Spectral variations among optically-classified L3 dwarfs, anchored to the L3 optical standard 2MASS~J1146+22 (red lines), from top to bottom: the 20-300~Myr young L3$\gamma$ G~196-3B, the young field L3$\beta$ 2MASS~J1726+15, the unusually red L3 dwarf 2MASS~J2151-24, the 3-5~Gyr companion G~62-33B, the unusually blue L3 2MASS~J1434+22, and the metal-poor sdL3.5 subdwarf SDSS~J1256-01. This sequence illustrates the range of gravity, cloud and metallicity variations that shape L dwarf NIR spectra. \label{f:l3}}
\end{figure}

As one of a few spectral libraries containing large numbers of UCDs, the SPL has enabled a broad range of stellar, brown dwarf and exoplanet science, and has been cited in over 100 publications to date. I highlight here three categories as examples:

\noindent {\bf Physical Properties of UCDs and Exoplanets:} The many atomic and molecular absorption features that characterize UCD NIR spectra are shaped by several factors, including photospheric temperature and elemental composition, pressure-sensitive opacity effects (e.g., collision-induced H$_2$ absorption, alkali line broadening; \citealt{2002A&A...390..779B,2003ApJ...583..985B}), and condensate formation and cloud properties \citep{2001ApJ...556..872A,2001ApJ...556..357A,2008ApJ...675L.105H}. These factors can significantly modify the NIR spectra of equivalently-classified UCDs (Figure~\ref{f:l3}), and several studies have used SPL data to disentangle these effects and extract the underlying physical properties of the sources.  Examples include surface gravity and age determinations for young M and L dwarfs \citep{2006ApJ...639.1120K,2010ApJ...715.1408Z,2013ApJ...772...79A,2013AJ....145....2F}; cloud characterization, particularly for unusually red and blue L dwarfs \citep{2008ApJ...674..451B,2008ApJ...686..528L}; metallicities of L subdwarfs \citep{2009ApJ...697..148B}; and determination of the masses and radii of individual T dwarfs \citep{2006ApJ...639.1095B,2007ApJ...655..522L,2010AJ....139..176F}.  SPL data have also been used to identify rare or benchmark UCD populations, including low-mass members of nearby moving groups and associations \citep{2007ApJ...669L..97L,2010A&A...515A..75A,2011ApJ...731....1S,2012ApJ...744..134M,2013ApJ...773...63A,2013MNRAS.430.1208P} and halo M, L and T subdwarfs \citep{2004ApJ...614L..73B,2010ApJ...708L.107L,2013arXiv1308.0495P}. Increasingly, the SPL has been used to analyze the spectra of directly-imaged exoplanets, such as HR~8799bcd, to characterize their physical and atmospheric properties \citep{2010ApJ...723..850B,2011ApJ...733...65B}.

\begin{figure}
\centerline{\includegraphics[height=5cm]{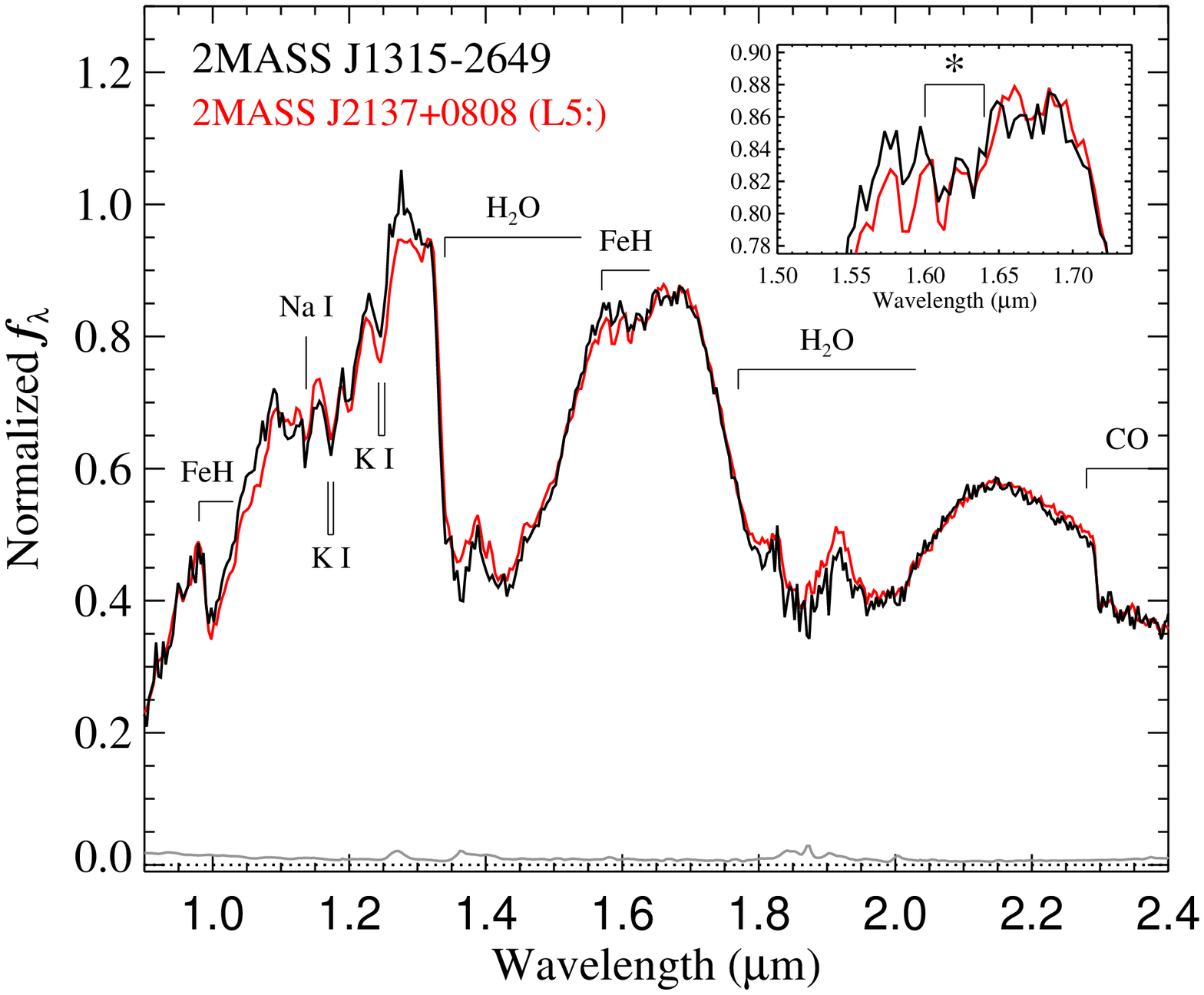}
\includegraphics[height=5cm]{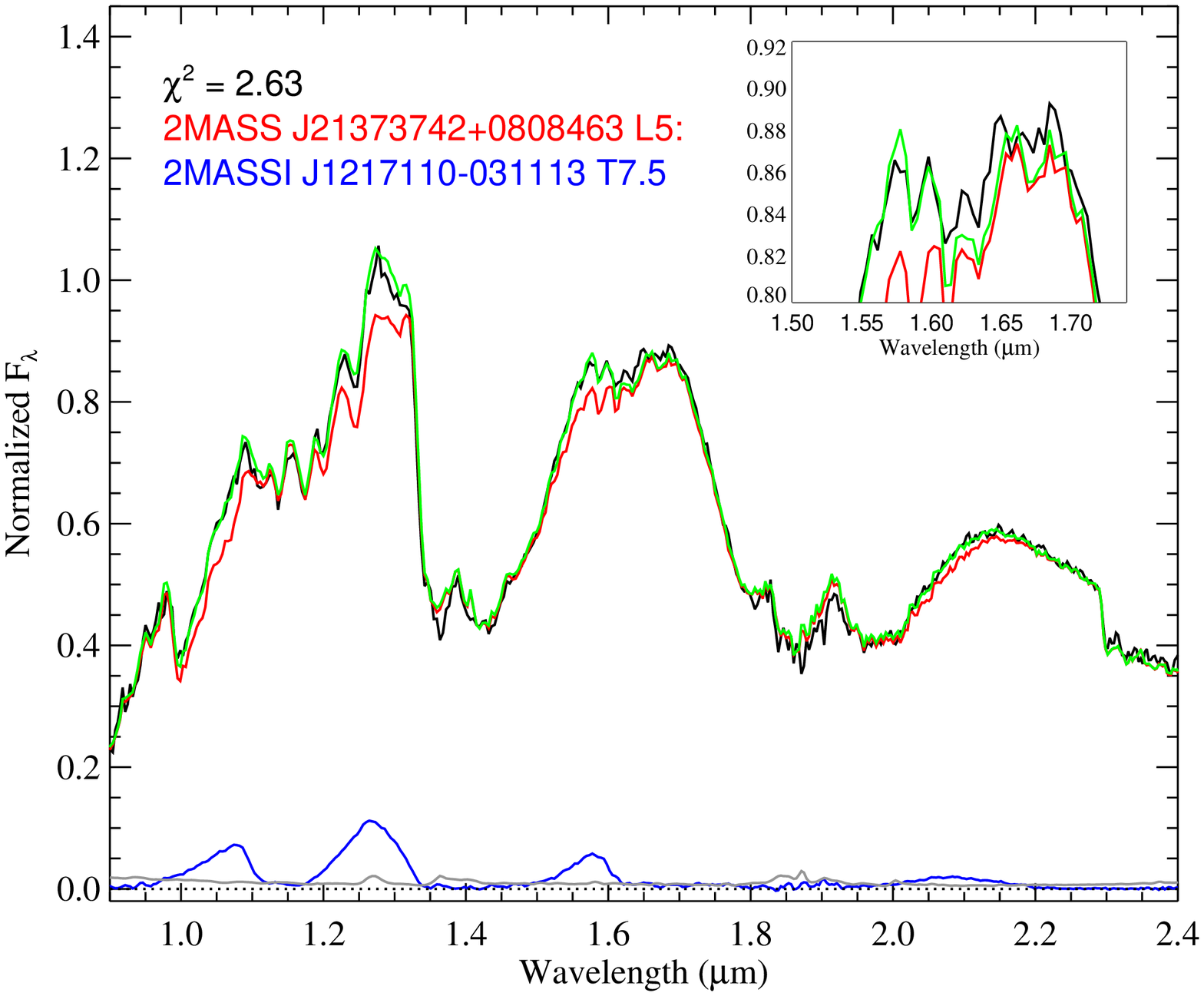}}
\caption{SPL analysis of the spectral binary 2MASS~J1315$-$2649 (black lines). The left panel shows best-fit single template, the L5 2MASS~J2137+0808 (red line), which has subtle deviations at 1.25~$\mu$m and 1.58~$\mu$m (inset box).  The right panel shows the best-fit binary template (green line) combining 2MASS~J2137+0808 (red line) with the T7.5 2MASS~J1217-0311 (blue line, scaled to relative fluxes), which is a statistically superior match. 2MASS~J1315$-$2649 was resolved as a 340~mas, $\Delta{K}$ = 5 binary, and resolved spectroscopy confirmed these classifications (from \citealt{2011ApJ...739...49B}) \label{f:sbinary}}
\end{figure}

\noindent {\bf Resolved and Unresolved UCD Binaries:} The substantial evolution of NIR spectral features between the M, L, and T classes implies that combined-light blends of multiples with different component spectral types can be distinctly peculiar.   To date, over 20 of these ``spectral binaries'' have been identified based on SPL data alone (e.g., \citealt{2004ApJ...604L..61C,2007AJ....134.1330B,2010ApJ...710.1142B,2010AJ....140..110G,2013ApJS..205....6M,2013MNRAS.430.1171D}).  Several of these systems have been resolved as extreme flux-ratio pairs (i.e., $\Delta{K} \approx 5$; \citealt{2011ApJ...739...49B}; Figure~\ref{f:sbinary}), short period radial velocity variables (P $<$ 1~yr; \citealt{2008ApJ...681..579B,2008ApJ...678L.125B,2012ApJ...757..110B}), astrometric variables \citep{2012ApJS..201...19D} and overluminous sources on color-magnitude diagrams \citep{2012ApJ...752...56F,2013arXiv1310.2191M}. SPL data have also been used to study resolved binaries, as templates to characterize components based on combined-light spectroscopy \citep{2006ApJS..166..585B,2006ApJ...639.1114R,2008ApJ...685.1183L,2009ApJ...697..824A,2012ApJS..201...19D} or resolved spectroscopy for widely-separated pairs (a $\gtrsim$ 1.5''; \citealt{2006AJ....132.2074M,2011AJ....141....7D,2013ApJ...772..129B}).

\noindent {\bf UCD Populations:} The SPL library has been a valuable resource for characterizing and simulating UCD populations, to design effective search (and rejection) strategies for wide-field imaging programs (e.g., \citealt{2010ApJ...709L..16O,2012ApJ...745..110O,2012ApJS..200...13B,2012ApJ...750...99T}); predict yields in deep surveys \citep{2009ApJ...695.1591P,2012ApJ...752L..14M};
and connect observable distributions in spectral type or effective temperature to the underlying substellar mass function, birth rate and multiplicity parameters \citep{2007ApJ...659..655B,2009ApJ...704.1519D}.

\subsection{Education and Outreach}\label{s:education}

The low resolution and compact format of SPL data makes it well-suited for education and outreach activities, particularly those aimed at teaching students and/or the general public about spectral classification and basic principles of radiative transfer.  A classification activity I frequently use for classroom settings is to divide 50 or so quarter-page prints of an assortment of  M, L and T dwarf spectra (including peculiar sources) among two or more teams, and then ask the students to organize the spectra into groups and sequences.  After they've converged on a sequence, they are asked to explain their choices, inspect the other team's classification (which they often find to be nearly identical both in ordering and grouping), and speculate on how their ordering relates to temperature, mass, etc. based on what they've learned about brown dwarfs.  More advanced students can develop their understanding of basic spectral analysis techniques by predicting NIR colors from the spectra, identifying specific molecular bands based on opacity charts, or specifying the spectral peculiarities associated with, e.g., young brown dwarfs, subdwarfs or binaries. These spectra also provide an excellent training set for classifying new discoveries as part of a student research project.

\subsection{Art}\label{s:art}

\begin{figure}
\centerline{\includegraphics[width=9cm]{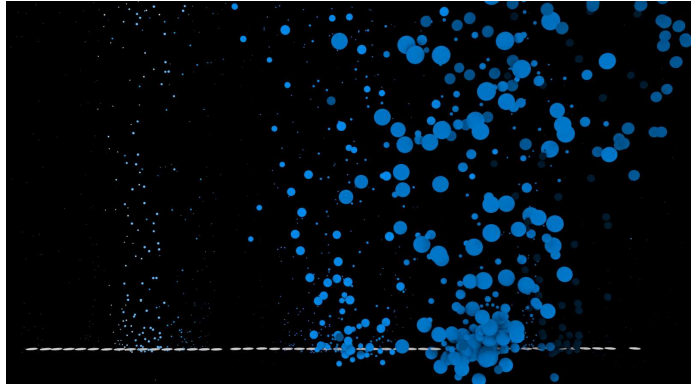}}
\caption{Still from the animation ``Spectral Bubblification'' created by UCSD undergraduate students Natasha Banchik and Ryan Phillips in 2013, based on SPL data for the T dwarf Gliese 570D.  Normalized flux is mapped to bubble density, wavelength is mapped to bubble size, and uncertainty is mapped to the projection of bubbles below the dashed line. This was one of several student art pieces generated from SPL data.  \label{f:art}}
\end{figure}

Finally, SPL data has been demonstrated as a source for data-driven art design. Figure~\ref{f:art} displays students' creative work from a data-driven art class I co-taught in 2013 at UC San Diego with Visual Arts faculty Michael Trigilio and Theatre Arts faculty Tara Knight. The piece uses SPL data for the T8 dwarf Gliese~570D to generate a bubbling animation, where the size, density and motion of the bubbles are determined by the data.  This is one of several visual and performance pieces created by the class, and illustrates the potential for astronomical libraries to create meaningful connections between scientists and artists.

\section{The SpeX Prism Library of the Future}\label{s:future}

Moving forward, I plan to expand the spectral contents of the SPL by integrating both published data from the community and publically-released, unpublished data from the forthcoming SpeX archive (A.\ Tokunaga, 2013, priv.~comm.). Uniform calibration will be assured by reprocessing all data acquired prior to 2007 (the most recent SpeXtool release), a project that will make use of a ``streamlined'' version of SpeXtool specifically developed for prism data. Meta-data for sources and spectra will also be expanded by adding 2MASS/DENIS/SDSS/WISE photometry and published astrometric and kinematic information. Online access of the data will be improved by integrating SQL query tools and basic visualization routines into the website. 
Data files will also be converted into VO-compliant formats according to the International Virtual Observatory Alliance (IVOA) Spectral Data Model \citep{2004IEEES..41h..29M}, allowing the use of online tools such as VOSpec \citep{2005ASPC..347..198O} and Starlink Spectral Analysis Tool \citep{2008ASPC..394..650C}. I also intend to release a python-based analysis toolkit ({\em splat}), currently under development, that is based on code developed out of the {\em astropy} project  \citep{2013A&A...558A..33A}.  

\begin{figure}
\centerline{\includegraphics[width=9cm]{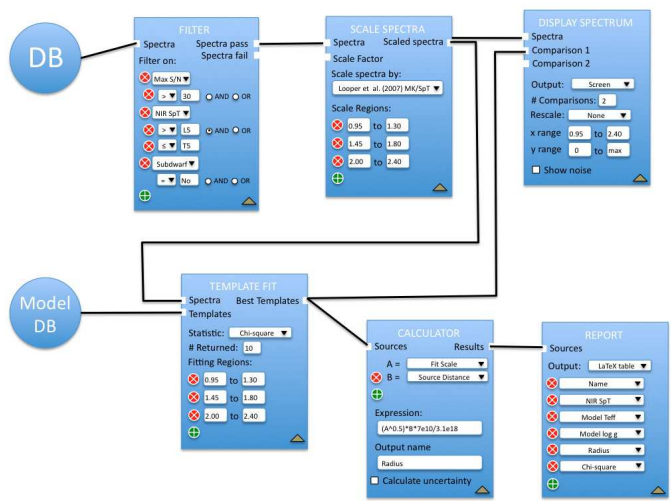}}
\caption{Illustration of a visual programming ``code'' to perform model fits of L5-T5 dwarfs in the SPL.  The top track filters the spectra by spectral type and data S/N, and scales them according to the M$_K$/spectral type relation of \citet{2008ApJ...685.1183L}. The bottom sequence draws from a library of model templates, finds the 10 best fits by minimizing $\chi^2$ in three spectral regions, and calculates the inferred radii for these sources.  The rightmost blocks display the spectra and model fits, and reports results to a LaTeX table. \label{f:vsplat}}
\end{figure}

Finally, in collaboration with computer science collaborators at UCSD, I am investigating the design of an online node-based visual programming tool that will enable non-experts to perform sophisticated spectral analyses of SPL data with minimal coding experience. Node-based architecture is widely used in instrument software development (e.g., LabVIEW), electronic media design (e.g., PD, Isadora), and programming tools for children (e.g., StarLogo, Lego Mindstorms), as it provides an easily-visualized dataflow model for algorithmic operations.  Figure~\ref{f:vsplat} displays a representation of a model-fitting analysis of SPL data with this architecture. The goal of this project is to facilitate the use of stellar spectral data analysis for pre-college project-based learning in astronomy and physics, and I plan to coordinate testing of this tool with middle and high school science teachers in the Imperial Unified School district, partnering with UCSD's Center for Research on Educational equity, Assessment \& TEaching (CREATE).

\section*{Acknowledgements}

The author extends his appreciation to those who have contributed to and made use of the SPL, and acknowledges the efforts of IRTF Director Alan Tokunaga, SpeX principal investigator John Rayner, 
SpeXtool developers Michael Cushing and Bill Vacca, and IRTF operators Dave Griep, Bill Golisch and Eric Volquardsen that have made SPL possible.
The author also thanks Daniella Bardalez Gagliuffi and Jason Morpeth for spot-checking and updating 
the SPL database.


\label{lastpage}
\end{document}